\documentclass[aip,apl,twocolumn,amsmath,amssymb,superscriptaddress]{revtex4}

\usepackage{array}
\usepackage{amsfonts}
\usepackage[dvips]{graphicx}
\usepackage{units}
\usepackage{dcolumn}
\usepackage{bm}
\usepackage{hyperref}
\usepackage{epstopdf}

\begin{document}

\title{Spin Seebeck Power Generators}
\author{Adam B. Cahaya}
\affiliation{Institute for Materials Research, Tohoku University, Sendai 980-8577, Japan}
\author{O.~A. Tretiakov}
\affiliation{Institute for Materials Research, Tohoku University, Sendai 980-8577, Japan}
\author{Gerrit E.~W. Bauer}
\affiliation{Institute for Materials Research and WPI-AIMR, Tohoku University, Sendai
980-8577, Japan}
\affiliation{Kavli Institute of NanoScience, TU Delft Lorentzweg 1, 2628 CJ Delft, The
Netherlands}
\date{\today }

\begin{abstract}
We derive expressions for the efficiency and figure of merit of two spin
caloritronic devices based on the spin Seebeck effect (SSE), i.e. the
generation of spin currents by a temperature gradient. The inverse spin Hall
effect is conventionally used to detect the SSE and offers advantages for
large area applications. We also propose a device that converts spin current
into electric one by means of a spin-valve detector, which scales favorably
to small sizes and approaches a figure of merit of 0.5 at room temperature.
\end{abstract}

\maketitle

Thermoelectric phenomena\cite{Sakata2005} transform heat currents into
electric power and vice versa. The Seebeck effect refers to the generation
of an electromotive force (emf) by a temperature gradient,\cite{Seebeck1823} while
the production of a heat current by an applied charge current is called
Peltier effect.\cite{Peltier} Thermoelectric power generators convert waste
heat into electric energy with many potential applications.\cite%
{Vining2009,Heremans2013} The spin degree of freedom adds functionalities
and may improve the efficiency of conventional thermoelectric devices.\cite%
{Bauer2012} The spin Seebeck effect (SSE) in the \textquotedblleft
longitudinal\textquotedblright\ configuration based on a ferromagnetic
insulator\cite{Uchida10} (FI) is especially promising. The SSE converts a
temperature difference between the FI and a normal metal (N) contact into
electric power\cite{Xiao10,Xiao10a} by pumping a spin current into the
normal metal that in turn is converted into a transverse emf by the inverse
spin Hall effect\cite{Hoffmann} (ISHE). The output power is proportional to
the device area perpendicular to the temperature gradient. This scaling
offers the opportunity to generate electricity by large-area coatings using
cheap materials.\cite{Kirihara12} Since here the paths of the charge and
heat currents are perpendicular to each other, alternative strategies to
enhance thermoelectric efficiency can be pursued.

In this Letter we validate the efficiency of SSE based power generators. In
addition to considering a device using the ISHE spin charge conversion, we propose harvesting electrical energy by a spin valve spin-filtering mechanism employing metallic ferromagnets. The spin-valve
based SSE power generator scales advantageously for the thermoelectric power
generation in small structures, and multiple elements can be easily added
for a higher output voltage, analogously to conventional thermopiles.

Thermoelectric generators produce electric power by the heat current that
flows between hot and cold reservoirs at temperatures $T_{H}$ and $T_{L},$
respectively. Its efficiency is a monotonic function of the dimensionless
figure of merit $ZT,$ where $T$ is the average temperature $\left(
T_{H}+T_{L}\right) /2$.\cite{Sakata2005} When $ZT\rightarrow \infty ,$ $\eta
\rightarrow \left( T_{H}-T_{L}\right) /T_{H}=\eta _{C}$, where $\eta _{C}$
is the maximum possible \textquotedblleft Carnot\textquotedblright\
efficiency. Here we derive $\eta \left( ZT\right) $ and $ZT$ for the two
types of generators driven by the SSE, taking into account the difference of
the physical mechanism between the SSE and conventional thermoelectrics.\cite%
{Bauer2012}

In the spin Seebeck effect, the spin current flowing through the interface
is caused by an imbalance of the spin pumping current due to magnetic
thermal noise $\mathbf{J}_{sp}$, that is proportional to FI's magnon temperature $T_{FI}^{m}$, and a fluctuating spin current caused by thermal
(Johnson-Nyquist) noise in the normal metal $\mathbf{J}_{fl}$ that is
proportional to N's electron temperature.\cite{Foros, Xiao09} Both currents
on average are polarized parallel to the magnetization direction ($\mathbf{%
\hat{m}}$) and cancel each other at equilibrium. The net SSE spin current
reads (indicating time average by $\left\langle \cdots \right\rangle $):\cite%
{Xiao10} 
\begin{equation}
J_{S}=\mathbf{\hat{m}}\cdot \left\langle \mathbf{J}_{sp}+\mathbf{J}%
_{fl}\right\rangle =L_{S}\left( T_{FI}^{m}-T_{N}^{{e}}\right) ,
\end{equation}%
where $L_{S}=\gamma \hbar G_{r}k_{B}/(2\pi eM_{s}V_{c})$ is the interfacial
response function, $\gamma $ is the gyromagnetic ratio, $G_{r}$ is the real
part of spin mixing conductance, $k_{B}$ is the Boltzmann constant, $M_{s}$
is the saturation magnetization, $e$ is the magnitude of electron's charge,
and $V_{c}$ is a magnetic coherence volume that depends on spin wave
stiffness and weakly on temperature.\cite{Xiao10a} Adachi \textit{et al.}%
\cite{Adachi11} and Hoffman \textit{et al}.\cite{Hoffman13} derived similar
expressions by different methods. The magnitude of the real part of the
mixing conductance $G_{r}$ is well established for intermetallic interfaces%
\cite{Brataas06} as well as interfaces with magnetic insulators.\cite{Jia11,
Burrowes12,weiler13}

\begin{figure}[tbp]
\includegraphics[scale=0.85]{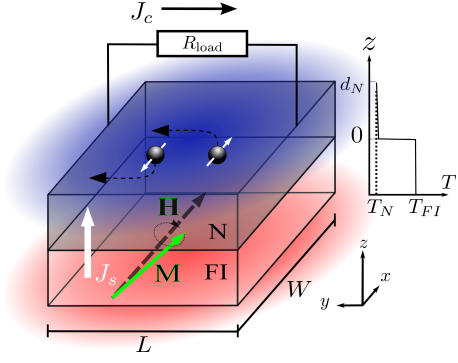}
\caption{(Color online) A schematic view of the spin Seebeck power generator based
on the inverse spin Hall effect (ISHE). A bilayer of ferromagnetic insulator and normal metal with a low interface heat conductance pumps a spin current $J_s$ into N. Then $J_s$ is converted into a transverse charge current $J_c$ by means of the ISHE. }
\label{ISHEsetup}
\end{figure}

The interface temperature discontinuity depends sensitively on the device
and material parameters.\cite{Schreier13} In the limit of small interface
heat resistance, the phonon temperature is continuous and governed by the
coupled heat diffusion equation for the bilayer with many not very
well-known parameters.\cite{Sanders77,Xiao10,Schreier13} Here we assume, for
simplicity, a dominating thermal boundary resistance $R_{K}=1/G_{K}$
(Kapitza resistance\cite{Kapitza41}), such that the magnon and phonon
temperatures in the FI at the interface $\left( z=0\right) $ are
approximately the same,\cite{Agrawal12, Schreier13} $T_{FI}^{m}(z=0^{-})%
\approx T_{FI}^{p}(z=0^{-})=T_{FI}$, see Fig.~\ref{ISHEsetup}. In this limit 
\begin{equation}
T_{FI}=T_{N}+R_{K}J_{Q},
\end{equation}%
where $T_{N}$ is the electron (and phonon) temperature of the normal metal
at $z=0^{+}$ and $J_{Q}$ is the heat current through the interface. Since
the spin contribution to the interface heat transport is comparably small,
the Kapitza resistance $R_{K}$ is dominated by phonon transport.\cite%
{Xiao10,Schreier13}

The heat flow through the system is partly converted into a spin current at
the interface that subsequently has to be transformed into electric energy. The
coupling between heat and spin over the FI$|$N interface can be written in
the form of a linear response matrix relation to the driving forces, \textit{viz}. the spin accumulation $\mu _{s}$ in the normal metal and temperature
difference $\Delta T=T_{N}-T_{FI},$ leading to the spin $J_{s}$ and averaged
heat $J_{Q}=\left( J_{Q}^{\mathrm{IN}}+J_{Q}^{\mathrm{OUT}}\right) /2
$ current responses: 
\begin{equation}
\left( 
\begin{array}{c}
J_{s} \\ 
J_{Q}%
\end{array}%
\right) =G_{S}\left( 
\begin{array}{cc}
1 & S_{S} \\ 
\Pi _{S} & G_{K}/G_{S}+S_{S}\Pi _{S}%
\end{array}%
\right) \left( 
\begin{array}{c}
-\mu _{s}/2e \\ 
-\Delta T%
\end{array}%
\right) .  \label{sse}
\end{equation}%
Here $G_{S}$ is the interface spin injection conductance, \cite{Bender,Jiao}
that generates backflow of the spin Seebeck spin current, 
$S_{S}=\left( \mu
_{s}/\left( 2e\Delta T\right) \right) _{J_{s}=0}=L_{S}/G_{S}$ is the spin
Seebeck coefficient, and $G_{K}=-\left( J_{Q}/\Delta T\right) _{J_{s}=0}$ is the Kapitza conductance (inverse of the Kapitza resistance $G_K=1/R_K$).
The spin current is positive when $\Delta T<0$ and $\mu _{s}<0$, so $%
G_{S},S_{S}>0$. The spin Peltier coefficient $\Pi _{S}=S_{S}T$ due to
Onsager reciprocity, where $T=\left( T_{N}+T_{FI}\right) /2$. 

We first consider the efficiency of a SSE generator with spin-charge
conversion by the ISHE when connected to an external load resistance $R_{%
\mathrm{load}}$ that utilizes the electric energy. The basic setup is shown
in Fig.~\ref{ISHEsetup}. Eq.~(\ref{sse}) defines the spin and heat currents
through the FI$|$N interface. The transverse electric current density $%
\mathbf{j}_{c}$ generated by the ISHE inside N at distance $z$ from interface is 
\begin{equation}
\mathbf{j}_{c}\left( z\right) =\theta _{\mathrm{SH}}j_{s}\left( z\right) 
\mathbf{\hat{z}}\times \mathbf{\hat{m}},  \label{current_ISHE}
\end{equation}%
where $j_{s}\left( z\right) \mathbf{\hat{z}}$ is the spin-current density
direction vector (in units of A/m$^{2}$), $\mathbf{\hat{m}}$ is its
polarization, and $\theta_\mathrm{SH}$ is the spin Hall angle. For $\mathbf{\hat{m}=\hat{x}}$ the charge current and emf are $%
\mathbf{j}_{c}=j_{c}\mathbf{\hat{y}\ }$and $\boldsymbol{\nabla }\mu =e%
\mathbf{E=\hat{y}}\partial _{y}\mu _{c}$. In the presence of spin flips, a
spin accumulation profile $\boldsymbol{\nabla }\mu _{s}=\mathbf{\hat{z}}%
\partial _{z}\mu _{s}$ builds up at the interface. It obeys the
spin-diffusion equation $\partial _{z}^{2}\mu _{s}=\mu _{s}/\lambda ^{2}$,
where $\lambda $ is the spin-flip diffusion length in N. The charge and spin current densities in N therefore read 
\begin{equation}
\left( 
\begin{array}{c}
{j}_{c} \\ 
{j}_{s} 
\end{array}%
\right) =-\sigma _{N}\left( 
\begin{array}{cc}
1 & \theta _{\mathrm{SH}} \\ 
-\theta _{\mathrm{SH}} & 1%
\end{array}%
\right) \left( 
\begin{array}{c}
\partial _{y}\mu _{c}/e \\ 
\partial _{z}\mu _{s}/\left( 2e\right)%
\end{array}%
\right) .  \label{SH}
\end{equation}%
Spin current conservation at the boundaries $z=0,d_{N}$ gives $%
j_{s}(z=0)=J_{s}/\left( WL\right) $ from Eq.~(\ref{sse}) and $j_{s}(d_{N})=0$%
, where $L$ is the length of N in the direction of the ISHE current, and $W$
is the width of the FI$|$N bilayer (see Fig.~\ref{ISHEsetup}). The solution
of the spin-diffusion equation 
\begin{eqnarray}
\frac{\mu _{s}}{2e} &=&\frac{\theta _{\mathrm{SH}}V\frac{\lambda }{L}\left(
G_{S}\sinh \frac{z}{\lambda }+G_{N}\cosh \frac{z}{\lambda }\right) }{%
G_{S}\cosh \frac{d_{N}}{\lambda }+G_{N}\sinh \frac{d_{N}}{\lambda }}  \notag
\\
&&-\frac{\left( \theta _{\mathrm{SH}}G_{N}V\frac{\lambda }{L}%
+G_{S}S_{S}\Delta T\right) \cosh \frac{d_{N}-z}{\lambda }}{G_{S}\cosh \frac{%
d_{N}}{\lambda }+G_{N}\sinh \frac{d_{N}}{\lambda }}
\end{eqnarray}%
depends on the spin conductance $G_{N}=\sigma _{N}WL/\lambda $ and the
induced transverse voltage $V=-L\partial _{y}\mu _{c}/e$. The integrated
transverse charge current $J_{c}$ in N then reads

\begin{eqnarray}
J_{c} &=&-\frac{G_{N}\lambda }{L}\left[ \frac{\theta_\mathrm{ISHE}%
G_{S}S_{S}\Delta T\tanh \frac{d_{N}}{2\lambda } }{G_{S}\coth \frac{d_{N}}{%
\lambda }+G_{N}} \right.  \notag \\
& &\left. +\frac{d_NV}{L} +\frac{\theta^2 _{\mathrm{ISHE}}\lambda V}{L} 
\frac{G_{S}+2G_{N} \tanh \frac{d_{N}}{2\lambda } }{G_{S}\coth \frac{d_{N}}{%
\lambda }+G_{N}} \right].
\end{eqnarray}

\begin{figure}[tbp]
\includegraphics[scale=0.75]{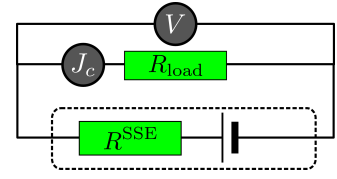}
\caption{(Color online) An effective circuit for the spin Seebeck power generators with internal
resistance $R^{\mathrm{SSE}}$. Current $J_c$ and voltage $V$ on external
load with $R_{\mathrm{load}}$ are measured by ideal volt and ampere meters. }
\label{equivcircuit}
\end{figure}

The SSE generator is a battery with internal resistance $R_{\mathrm{ISHE}}^{%
\mathrm{SSE}}\propto L/W$ and a maximum output voltage $V_{\mathrm{ISHE}}^{%
\mathrm{SSE}}$, \cite{Cahaya2013} see Fig.~\ref{equivcircuit}, that is 
\begin{align}
J_{c}& =\frac{V_{\mathrm{ISHE}}^{\mathrm{SSE}}-V}{R_{\mathrm{ISHE}}^{\mathrm{%
SSE}}},  \label{J_battery} \\
V_{\mathrm{ISHE}}^{\mathrm{SSE}}& =-R_{\mathrm{ISHE}}^{\mathrm{SSE}}\frac{%
\theta _{\mathrm{SH}}\sigma _{N}WG_{S}S_{S}\Delta T\tanh \frac{d_{N}}{%
2\lambda }}{G_{S}\coth \frac{d_{N}}{\lambda }+G_{N}}\propto L, \\
\frac{1}{R_{\mathrm{ISHE}}^{\mathrm{SSE}}}& =\frac{\sigma _{N}W\lambda }{L}%
\left( \frac{d_{N}}{\lambda }+\theta _{\mathrm{SH}}^{2}\frac{%
G_{S}+2G_{N}\tanh \frac{d_{N}}{2\lambda }}{G_{S}\coth \frac{d_{N}}{\lambda }%
+G_{N}}\right) .
\end{align}%
The voltage drop over the load resistance $R_{\mathrm{load}}$ is 
\begin{equation}
V=\frac{R_{\mathrm{load}}}{R_{\mathrm{load}}+R_{\mathrm{ISHE}}^{\mathrm{SSE}}%
}V_{\mathrm{ISHE}}^{\mathrm{SSE}}\propto L.
\end{equation}%
The thermally generated electric power $\mathcal{W}=J_{Q}^{\mathrm{IN}%
}-J_{Q}^{\mathrm{OUT}}$ dissipated in the load resistance 
\begin{equation}
\mathcal{W}=\frac{V^{2}}{R_{\mathrm{load}}}=R_{\mathrm{load}}\left( \frac{V_{%
\mathrm{ISHE}}^{\mathrm{SSE}}}{R_{\mathrm{load}}+R_{\mathrm{ISHE}}^{\mathrm{%
SSE}}}\right) ^{2}
\end{equation}%
scales with the device area $WL$. The maximum output voltage $V_{\mathrm{ISHE%
}}^{\mathrm{SSE}}$ when $R_{\mathrm{load}}\rightarrow \infty $ is
proportional to the sample length $L$. $V_{\mathrm{ISHE}}^{\mathrm{SSE}}\sim
1/d_{N}$ for large $d_{N},$ since the emf is short-circuited by the
non-active conducting region. When $d_{N}\ll \lambda ,$ the voltage output
vanishes with the gradient of the spin accumulation.

The efficiency $\eta ^{\mathrm{SSE}}=\mathcal{W}/\left\vert J_{Q}\right\vert 
$ can be expressed in a form similar to conventional thermoelectrics, \cite%
{Sakata2005} 
\begin{equation}
\eta ^{\mathrm{SSE}}=\frac{\left\vert \Delta T\right\vert }{T_H}\frac{\sqrt{%
1+\left( ZT\right) ^{\mathrm{SSE}}}-1}{\sqrt{1+\left( ZT\right) ^{\mathrm{SSE%
}}}+1-\left\vert \Delta T\right\vert /T_H}  \label{efficiency}
\end{equation}%
in terms of figure of merit $\left( ZT\right) ^{\mathrm{SSE}}$ that can be obtained by maximizing $\eta^{\mathrm{SSE}}$ with respect
to $R_{\mathrm{load}}$, i.e. by impedance matching, leading to 
\begin{equation}
\left( ZT\right) _{\mathrm{ISHE}}^{\mathrm{SSE}}=\frac{4\theta _{\mathrm{SH}%
}^{2}e^{-d_{N}^{\mathrm{opt}}/{\lambda }}G_{N}S_{S}^{2}T/G_{K}}{\left( 1+%
\frac{G_{N}}{G_{S}}\right) \left( 1+\frac{G_{N}}{G_{S}}+\frac{G_{N}S_{S}^{2}T%
}{G_{K}}\right) }.
\end{equation}%
Here the thickness $d_{N}^{\mathrm{opt}}$ for the optimal spin-charge
conversion is given by the (positive) solution of the equation 
\begin{equation}
\frac{d_{N}^{\mathrm{opt}}}{\lambda }+\theta _{\mathrm{SH}}^{2}=\frac{1}{2}%
\sinh \frac{d_{N}^{\mathrm{opt}}}{\lambda }.
\end{equation}%
We can now estimate the figure of merit for the yttrium iron garnet (YIG)$|$Pt system $\left(
ZT\right) _{\mathrm{ISHE}}^{\mathrm{SSE}}$. The spin Seebeck coefficient has
the universal value $S_{S}\approx k_{B}/\left( 1.3\times e\right) \approx
65\ \mathrm{\mu VK}^{-1}.$\cite{Jiao} With\cite{weiler13} $G_{r}/A=\left(
e^{2}/h\right) 10^{19}\,\mathrm{m}^{-2}=4\times 10^{14}\Omega ^{-1}$\textrm{m%
}$^{-2}$ we find (here and in the following at room temperature) $%
L_{S}/A\sim 4\times 10^{9}$ AK$^{-1}$m$^{-2}$. The spin conductance $%
G_{S}=L_{S}/S_{S}$ governs the spin current injected back into FI by the
spin accumulation and is estimated for the YIG$|$Pt interface to be $%
G_{S}/A\sim 6\times 10^{13}\ \mathrm{\Omega }^{-1}\mathrm{m}^{-2},$ i.e.
much smaller than the spin conductance of Pt: $G_{N}/A=10^{15}$ $\Omega ^{-1}$%
\textrm{m}$^{-2}\ $for\cite{weiler13} $\lambda =1.5$ $\unit{nm}$ and $\rho
_{Pt}=500\ \mathrm{n\Omega m.}$ Limit $G_{S}/G_{N}\rightarrow 0$ leads to the
simplification%
\begin{equation}
\left( ZT\right) _{\mathrm{ISHE}}^{\mathrm{SSE}}\rightarrow 4\theta _{%
\mathrm{SH}}^{2}e^{-d_{N}^{\mathrm{opt}}/{\lambda }}\frac{L_{S}^{2}T}{G_{N}}%
\frac{1}{G_{K}+G_{S}S_{S}^{2}T}.
\end{equation}%
It shows that $\left( ZT\right) _{\mathrm{ISHE}}^{\mathrm{SSE}}$ is
invariant with respect to the sample area. We can estimate its value using%
\cite{weiler13} $\lambda =1.5$ $\unit{nm}$ and $\theta _{\mathrm{SH}}=0.1,$
as well as the phonon contribution to the Kapitza conductance\cite%
{Schreier13} $G_{K}^{(\mathrm{ph})}/A=1.6\times 10^{8}$\thinspace Wm$^{{-2}}$%
K$^{-1},$ which corresponds to about 40 nm of bulk YIG and is larger than
the spin contribution\cite{Schreier13} $G_{K}^{(\mathrm{m})}/A\sim 0.5\times
10^{8}$ $\text{Wm}^{-2}\text{K}^{-1}$. This leads to $L_{S}^{2}T/\left(
G_{N}G_{K}\right) \sim 0.025$ and $\left( ZT\right) _{\mathrm{ISHE}}^{%
\mathrm{SSE}}\sim 10^{-4}$ for the optimum width of $d_{N}^{\mathrm{opt}%
}=2.16\,\lambda $.

\begin{figure}[tbp]
\centering
\includegraphics[scale=0.85]{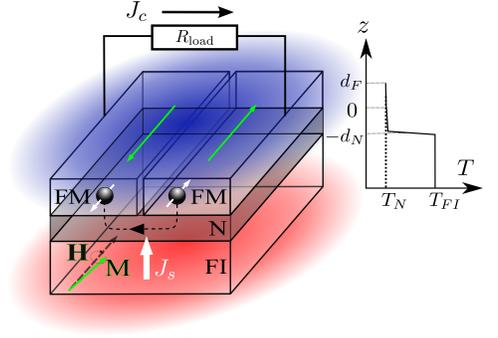}
\caption{(Color online) A schematic view of the spin-valve based spin Seebeck power
generator. Two antiparallel ferromagnetic layers (FM) are added to the FI$|$N bilayer device in Fig.~\ref{ISHEsetup}. The spin accumulation in N drives a charge current through the metallic spin valve, thereby generating a voltage over the load resistance. }
\label{SVsetup}
\end{figure}

We now turn to the alternative SSE power generator in which the thermal
spin-motive force generates an electromotive force by means of ferromagnetic
metal (FM) contacts, \textit{viz}. by the spin valve effect. The N layer is
now a metal with a long spin-flip diffusion length such as Cu. As shown in
Fig.~\ref{SVsetup}, the spin current in this case is injected into the N
layer of a lateral metallic spin valve with ferromagnetic contacts in an
antiparallel configuration collinear to the magnetization of the magnetic
insulator. The spin accumulation injected thermally into the N spacer
therefore generates a voltage difference between the contacts. The FM$|$N
contact areas are $A$ and that of the FI$|$N contact is approximately $2A$.

We assume that the spin-flip diffusion length is larger than the sample
dimensions such that the spin accumulation is constant in N. The charge-spin
linear response relations at the interface of the N$|$FM interface with
conductance $G_{I}$ (including the magnetically active thickness of the bulk
ferromagnet) can be written as 
\begin{equation}
\left( 
\begin{array}{c}
{j}_{c}^{\left( m\right) } \\ 
{j}_{s}^{\left( m\right) }%
\end{array}%
\right) =\frac{G_{I}}{eA}\left( 
\begin{array}{cc}
1 & mP \\ 
mP & 1%
\end{array}%
\right) \left( 
\begin{array}{c}
\mu ^{\left( m\right) }(0)-\mu ^{N} \\ 
\left( \mu _{s}^{\left( m\right) }(0)-\mu _{s}^{N}\right) /2%
\end{array}%
\right) ,
\end{equation}%
where $P\equiv (G_I^\uparrow-G_I^\downarrow)/(G_I^\uparrow+G_I^\downarrow)$ is the spin polarization of the N$|$FM contact, $\mu ^{N}$ and $\mu _{s}^{N}$ are the electrochemical potential and
spin accumulation in N, $\mu ^{(m)}$ and $\mu _{s}^{(m)}$ are the electrochemical potential and
spin accumulation in FM, and the superscript $m=\pm $ corresponds to a
contact with magnetization parallel ($+$) or antiparallel ($-$) to that of
the FI.

Assuming zero spin accumulation on the other side of the FM contact and
conservation of spin current in N, $J_{s}^{N}=J_{s}^{0}+\sum_{\pm
}J_{s}^{\pm }(0)=0$: 
\begin{align}
\frac{\mu _{s}^{N}}{2e}=& \frac{-G_{S}S_{S}\Delta T-PG_{I}V}{G_{S}+2G_{I}},
\\
J_{c}=& \frac{PS_{S}\Delta T/G_{I}}{1+2G_{I}/G_{S}}-\frac{G_{I}V}{2}\frac{%
1+2(1-P^{2})G_{I}/G_{S}}{1+2G_{I}/G_{S}},
\end{align}%
where the induced voltage $V=(\mu ^{-}-\mu ^{+})/e$. The effective electric
circuit for the spin-valve based generator is again given by Fig. \ref%
{equivcircuit} with internal resistance $R_{\mathrm{SV}}^{\mathrm{SSE}}$ and
maximum voltage $V_{\mathrm{SV}}^{\mathrm{SSE}}$: 
\begin{align}
V_{\mathrm{SV}}^{\mathrm{SSE}}& =\frac{2PS_{S}\Delta T}{%
1+2(1-P^{2})G_{I}/G_{S}}, \\
R_{\mathrm{SV}}^{\mathrm{SSE}}& =\frac{2}{G_{I}}\frac{1+2G_{I}/G_{S}}{%
1+2(1-P^{2})G_{I}/G_{S}}.
\end{align}%
The optimal figure of merit $\left( ZT\right) _{\mathrm{SV}}^{\mathrm{SSE}}$
is obtained again by maximizing the efficiency $\eta ^{\mathrm{SSE}}$ with
respect to $R_{\mathrm{load}}$: 
\begin{equation}
\left( ZT\right) _{\mathrm{SV}}^{\mathrm{SSE}}=\frac{%
2P^{2}G_{I}S_{S}^{2}T/G_{K}}{\left( 1+\frac{2G_{I}}{G_{S}}\right) \left[
1+2(1-P^{2})\left( \frac{G_{I}}{G_{S}}+\frac{G_{I}S_{S}^{2}T}{G_{K}}\right) %
\right] }.
\end{equation}%
For an intermetallic interface, $G_{S}\ll G_{I}$ \cite{Hamrle05} and 
\begin{equation}
\lim_{G_{S}\rightarrow 0}\left( ZT\right) _{\mathrm{SV}}^{\mathrm{SSE}}=%
\frac{P^{2}}{1-P^{2}}\frac{G_{S}}{2G_{I}}\frac{G_{S}S_{S}^{2}T}{G_{K}}.
\end{equation}%
In the limit of a half-metal, this expression appears to diverge, but when
we first take $P\rightarrow 1$ and then $G_{S}/G_{I}\rightarrow 0$:
\begin{equation}
\lim_{G_{S}/G_{I}\rightarrow 0}\lim_{P\rightarrow 1}\left( ZT\right) _{%
\mathrm{SV}}^{\mathrm{SSE}}=\frac{G_{S}S_{S}^{2}T}{G_{K}}\sim 0.5,
\end{equation}%
the result looks similar to the figure of merit for conventional thermoelectrics.
The numerical estimate is obtained for\cite{Schreier13} $G_{K}/A\sim
1.6\times 10^{8}$ m$^{-2}$K$^{-1}$ and\cite{weiler13} $G_{r}/A\sim 4\times
10^{14}$ $\Omega ^{-1}$m$^{-2}$.

To summarize, we consider two schemes of thermoelectric power
generators based on the spin Seebeck effect. We estimate their figures
of merit $\left( ZT\right) ^{\mathrm{SSE}}$ under the assumption that the total heat
conductance is limited by the FI$|$N interface. 
This assumption importantly simplifies the model, but since the Kapitza interface conductance has not yet been measured for YIG$|$metal interfaces, also introduces uncertainties. The output voltage of the
SSE-ISHE device is proportional to sample length $L$ perpendicular to the
FI's magnetization and temperature gradient (Fig. \ref{ISHEsetup}), while
the power scales with the area. Therefore this scheme has an advantage for
large area devices, but $\left( ZT\right) _{\mathrm{ISHE}}^{\mathrm{SSE}}$
is small since it is limited by $\theta _{\mathrm{SH}}^{2}\exp \left(
-d_{N}/\lambda \right) $. A spin valve can convert spin into charge current
as well, offering the possibility to enhance $ZT$
considerably. The scale independence of the output voltage in spin-valve SSE
devices can be useful for micro- and nanoscale applications, since the
output voltage does not decrease when down-scaling the device. Experiments
demonstrating the SSE by a spin valve are highly desirable since they would
shed light on the role of interface proximity or spin-orbit interaction
effects that might exist for YIG$|$Pt but not for YIG$|$Cu.\cite{Huang} The present modelling is also applicable for other devices, such as spin Seebeck-assisted magnetic random access memories.\cite{Mojum} 

We thank E. Saitoh, K. Uchida, J. Flipse, and J. Xiao for insightful
discussions. A.B.C. is supported by Japanese Ministry of Education Culture,
Sports, Science and Technology (MEXT) Scholarship Grant. We acknowledge
support by the Grants-in-Aid for Scientific Research (Nos. 25800184,  25220910, and
25247056), the DFG via SPP 1538 \textquotedblleft Spin Caloric
Transport\textquotedblright , the EU RTN Spinicur and DAAD SpinNet.

\end{document}